\documentclass[a4paper,fleqn,usenatbib,letters]{mnras}

\usepackage[T1]{fontenc}
\usepackage{ae,aecompl}

\usepackage{graphicx}	
\usepackage{amsmath}	
\usepackage{amssymb}	
\usepackage{latexsym}
\usepackage{url}

\def\mnras{MNRAS}
\def\apj{ApJ}
\def\aap{A\&A}
\def\apjl{ApJL}
\def\apjs{ApJS}

\def\ga{\sim}

\title[Photoevaporation of TW~Hya]{A photo-evaporative gap in the closest planet forming disc}

\author[Ercolano, Rosotti, Picogna, Testi]{Barbara Ercolano$^{1,2}$\thanks{E-mail: ercolano@usm.lmu.de (BE)}, Giovanni P. Rosotti$^{3}$,  Giovanni Picogna$^{1}$, Leonardo Testi$^{2,4,5}$\\
$^{1}$Universit\"ats-Sternwarte M\"unchen, Scheinerstr. 1, 81679 M\"unchen, Germany\\
$^{2}$Excellence Cluster Origin and Structure of the Universe,
Boltzmannstr.2, 85748 Garching bei M\"unchen, Germany\\
$^3$Institute of Astronomy, University of Cambridge, Madingley Road, Cambridge, UK\\
$^4$European Southern Observatory, Garching bei M\"unchen, Germany\\
$^5$INAF/Osservatorio Astrofisico di Arcetri, Largo E. Fermi 5, I-50125 Firenze, Italy}

\date{Accepted XXX. Received YYY; in original form ZZZ}

\pubyear{2016}

\begin{document}

\label{firstpage}

\pagerange{\pageref{firstpage}--\pageref{lastpage}}

\maketitle

\begin{abstract}

The dispersal of the circumstellar discs of dust and gas surrounding young low-mass stars has important implications for the formation of planetary systems. Photoevaporation from energetic radiation from the central object is thought to drive the dispersal in the majority of discs, by creating a gap which disconnects the outer from the inner regions of the disc and then disperses the outer disc from the inside-out, while the inner disc keeps draining viscously onto the star. In this Letter we show that
the disc around TW Hya, the closest protoplanetary disc to Earth, may be the first
object where a photoevaporative gap has been imaged around the time at which it is being created. Indeed the detected gap in the ALMA images is consistent with the expectations of X-ray photoevaporation models, thus not requiring the presence of a planet. The photoevaporation model is also consistent with a broad range of properties of the TW Hya system, e.g.  accretion rate and the location of the gap at the onset of dispersal.
We show that the central, unresolved $870\ \mu\mbox{m}$ continuum source might be produced by free free emission from the gas and/or residual dust inside the gap. 

\end{abstract}

\begin{keywords}
protoplanetary discs
\end{keywords}

\section{Introduction}

Planets form from the reservoir of dust and gas residing in the circumstellar discs which surround young stars.
The physical and chemical characteristics of this material influence the process of planet formation and are in turn influenced by the energetic radiation from the young stellar object.
Indeed, young stars are strong sources of X-ray radiation, with approximately one thousandth of their bolometric luminosity being emitted at energies higher than $100\ \mbox{eV}$.
X-rays play an important role in the disc evolution, by providing ionisation over a large range of columns (e.g. Igea \& Glassgold 1999; Ercolano \& Glassgold 2013), which is important for the chemical evolution of the material as well as for angular momentum transport mechanisms.
In particular, ionisation of the disc atmosphere by the ``soft X-rays'' ($100\ \mbox{eV}<\mbox{eV}<1\ \mbox{keV}$) heats the gas and launches a thermal wind (e.g. Owen et al. 2010), which is able to disperse the disc, via (a) the formation of a gap, followed by (b) a hole which quickly grows in radius, with the outer disc being eroded from the inside-out.

The X-ray photoevaporation model (Ercolano et al. 2008, 2009; Owen et al. 2010, 2011, 2012) is successful in reproducing the observed two-timescale dispersal (e.g. Luhman et al. 2010; Koepferl et al. 2013), and it can reproduce the intensities and profiles of emission lines produced in the wind (Ercolano \& Owen, 2010, 2016).
However, the direct observation of a disc undergoing gap formation via photoevaporation is still lacking.
In this Letter we propose that the gap at $1\ \mbox{au}$ in the nearest protoplanetary disc, TW~Hya, observed in the $870\ \mu\mbox{m}$ continuum by Andrews et al. (2016), using the long-baseline Atacama Large Millimeter/submillimeter array (ALMA), may indeed be the first directly imaged photoevaporative gap in a solar-type star to date.
Surprisingly, this possibility has not yet been properly explored in the recent literature.

This Letter is organised as follows: In Section 2 we summarise the relevant characteristics and current evidence for photoevaporation in the TW~Hya system. In Section 3 we explore the possible origin of the unresolved $870\ \mu\mbox{m}$ emission from the inner $0.5\ \mbox{au}$ of the disc. In Section 4 we show dust and gas density distributions from a numerical model of a 1D viscously evolving disc, subject to X-ray photoevaporation.
Our conclusions are summarised in Section 5.

\section{The TW~Hya system: what is known about our closest neighbour }

TW~Hya is considered a young solar analogue with a central stellar mass of approximately $0.7\ M_\odot$.
Being the closest gas-rich protoplanetary disc ($54\ \mbox{pc}$, van Leeuwen 2007), this object has been subject of a number of observational campaigns dating back from the early nineties (e.g. Blondel et al. 1993).
Its nearly pole-on inclination ($i \sim 7^o$, Qi et al. 2004) further contributes to making TW~Hya one of the reference objects for the study of protoplanetary discs.

\subsection{Disc structure}

Calvet et al. (2002) first reported evidence for a dust-depleted inner hole, which they placed at a radius of $4\ \mbox{au}$ via detailed radiative transfer modelling of its spectral energy distribution (SED).
On the basis of the $10\ \mu\mbox{m}$ silicon emission, Calvet et al (2002) concluded that the inner region cannot be completely devoid of dust, and they estimate that $\sim 0.5$ lunar mass of $\sim 1\ \mu\mbox{m}$ particles must be still present inside the cavity, and interpret this as evidence of a {\it developing} gap in the disc.
The exact size of the gap of TW~Hya has been an object of dispute in the literature.
Near-infrared (Eisner et al. 2006) and mm (Hughes et al. 2007) interferometry confirmed a size of $4\ \mbox{au}$ for the central hole, while a gap radius of only $0.7\ \mbox{au}$ was obtained by Ratzka et al. (2007) via mid-infrared intereferometric observations.
The observations can however be reconciled if instead of an inner hole completely devoid of dust, an additional optically thick ring with a $0.5\ \mbox{au}$ radius is considered (e.g. Akeson et al. 2011).
More recently Menu et al (2014) presented a new model, which was able to fit all interferometric observations from near-infrared to \mbox{cm} wavelengths available at the time.
This model required an inner cavity with radius $0.3-0.5\ \mbox{au}$ with a maximum in the surface density occurring at $\sim 2.5\ \mbox{au}$.

While the presence of a gap or at least of a \textit{dust} depleted region in the inner disc has been confirmed by all interferometric observations, the detection of a non-negligible accretion signature (4$\times 10^{-10}\ M_{\odot}/\mbox{yr}$ to $\sim 10^{-9}\ M_{\odot}/\mbox{yr}$, Muzerolle et al., 2000; Eisner et al. 2010; Manara et al. 2014), implies that a significant amount of \textit{gas} must still be present inside the dust cavity.

\subsection{Gaps and rings: the ALMA view}

Andrews et al. (2016) presented long-baseline ALMA observations of the $870\ \mu\mbox{m}$ continuum from TW~Hya, which are sensitive to millimeter-sized dust grains down to a spatial scale of $1\ \mbox{au}$.
The data show a series of concentric dark and bright shallow ring-like structures in the outer disc, while the inner disk is dominated by a dark annulus centred at $1\ \mbox{au}$, separating a bright ring that peaks at $2.4\ \mbox{au}$ and an unresolved continuum source less than $0.5\ \mbox{au}$ in radius and emitting $\sim 1\ \mbox{mJy}$ in flux.
The modest brightness contrast of some of the concentric features at r $> 10\ \mbox{au}$ may be associated with condensation fronts of abundant volatiles (Cuzzi \& Zahnle, 2004; Okuzumi et al. 2015), or may be evidence of zonal flows (Johansen et al. 2009).

The dark annulus at $\sim 1\ \mbox{au}$ must have, however, a different origin.
The ALMA observations together with the dip in the SED at mid-infrared wavelengths (Calvet et al. 2002) suggest that the gap must be depleted of grains from $\mu\mbox{m}$ to $\mbox{cm}$ size.
On that basis, Andrews et al. (2016) argue against the dip being a consequence of changing solid properties at around the H$_2$O snow-line.
As an alternative, they suggest that a more detailed modeling of the ALMA data may show whether a young super-Earth may be responsible for dynamically carving the observed gap (e.g., Picogna \& Kley 2015, Rosotti et al. 2016). Surprisingly, the photoevaporation scenario is not considered at all in that work.

\vspace{-0.4cm}
\subsection{Evidence of X-ray photoevaporation in TW~Hya }

The TW~Hya central object is known to emit substantial X-ray radiation, with a total X-ray flux ($0.1\ \mbox{keV}<E<10\ \mbox{keV}$) of approximately $2.5\times10^{30}\ \mbox{erg/sec}$, based on an extrapolation to lower and higher energies of the $0.3-5\ \mbox{keV}$ XMM-Newton spectrum (Robrade \& Schmitt, 2006).
Following Equation 9 of Owen, Clarke \& Ercolano (2012), the corresponding mass loss rate due to X-ray photoevaporation is $\sim 2\times10^{-8}\ M_{\odot}/\mbox{yr}$,
which is roughly a factor of $10$ higher than the measured accretion rate for TW Hya.  Owen et al (2010) predict that for a $0.7\ \mbox{M}_\odot$ star and X-ray luminosity of $2\times10^{30}\ \mbox{erg/sec}$ X-ray photoevaporation will create a gap at  $\sim 1\ \mbox{au}$, when the accretion rate falls to about a tenth of the wind rate.  (e.g. see their Figure 11).
Hence, the dark annulus detected by Andrews et al. (2016) may indeed be the first stage of the dispersal phase of the TW Hya disc.

Direct evidence of currently ongoing photoevaporation is provided by the detection of a few $\mbox{km/s}$ blueshift in the profile of the [NeII]~$12.8\ \mu\mbox{m}$ line (Herczeg et al. 2007; Pascucci \& Sterzik 2009), which is well fit by both EUV and X-ray photoevapoaration models (Alexander 2008, Ercolano \& Owen, 2010).
The [OI]$6300$ has also been shown to be produced in a slow-moving quasi-neutral X-ray photoevaporative wind (Ercolano \& Owen, 2016), and is often observed to be blue-shifted with a low-velocity component of a few $\mbox{km/s}$.
In the case of TW~Hya the [OI]$6300$ was found to be centred on the stello-centric velocity (Pascucci et al. 2011).
This is not surprising given TW~Hya near face-on inclination and the fact that the [OI]$6300$ line is emitted from the very inner regions of the disc wind (Ercolano \& Owen 2010, 2016), where the underlying disc is devoid of dust, allowing us to see the redshifted outflowing gas.
The $10\ \mbox{km/s}$ FWHM and $\sim10^{-5}\ \mbox{L}_\odot$ line luminosity measured by Pascucci et al. (2011) for [OI]$6300$ are also well reproduced by the X-ray photoevaporation models presented by Ercolano \& Owen (2016) for accretion and X-ray luminosities and the inclination of TW~Hya.

\section{Origin of the unresolved $870\ \mu\mbox{m}$ emission from the inner $0.5\ \mbox{au}$}
Andrews et al. (2016) report $\sim 1\ \mbox{mJy}$ flux at $870\ \mu\mbox{m}$ as an unresolved continuum source from the inner $\sim 0.5\ \mbox{au}$ of the TW~Hya disc.
This might look incompatible with a photo-evaporative origin for the gap. Indeed, Alexander \& Armitage (2007) found that large grains disappear quickly ($10^4$ yr) from the inner disc due to radial drift, suggesting that a gap would turn quickly into a hole, with no detectable central point source. In this section we show, however, how free-free gas emission or residual dust inside the gap can easily explain the observed flux, supporting the photo-evaporation hypothesis.

\subsection{Free-free emission}

Pascucci et al. (2012) analysed the long wavelength portion of the SED from TW~Hya covering from $870\ \mu\mbox{m}$ out to 6~cm (data from Wilner et al. 2005 and Andrews et al. 2012) and came to the conclusion that about half of the flux detected at 3.5cm could not be accounted for by dust emission. They attributed the extra 140$\pm$40$\mu$Jy at 3.5$\mu$m to a photoevaporative disc wind. However, emission from a disc wind is expected to originate from a region far more extended than $0.5\ \mbox{au}$, and thus it cannot explain the unresolved $870\ \mu\mbox{m}$ emission detected by Andrews et al. (2016).  Dense ionised gas, however, exists also in the form of a stellar wind or an ionised bound inner disc atmosphere.
One can obtain an upper limit to the contribution from gas free-free emission to the sub-mm flux by assuming a spherical optically thick wind. In this case the SED emission in the radio can be described by a power law with index 0.6 (Panagia \& Felli, 1975)\footnote{Steeper indices are obtained by e.g. Franco et al. (2000), but the latter are unlikely in this case and more appropriate to hypercompact HII regions.}. A 0.93mJy flux at $870\ \mu\mbox{m}$ would then correspond to $\sim 100\ \mu\mbox{Jy}$ at $3.5\ \mbox{cm}$, explaining thus most of the 140$\pm$40$\mu$Jy excess flux at 3.5cm reported by Pascucci et al. (2012).  This would lead to the conclusion that the compact, unresolved source in Andrews et al. (2016) can be caused by a stellar wind and there is no need to invoke the presence of dust in the inner disk.
Conversely, this is an upper limit to the emission of the wind at mm-wavelengths: if the wind is optically thin, the spectral index is also expected to be lower, meaning that its contribution in the (sub-)mm will be lower. It is thus not excluded that a fraction of the compact unresolved emission at 870$\mu$m may be caused by dust.

We have calculated photoionisation models of the draining inner disc of TW Hya with {\sc mocassin} (Ercolano et al. 2003, 2005, 2008b) using the the inner $0.5\ \mbox{au}$ of the gas and dust density distributions from Ercolano \& Owen (2016). This consists of a hydrodynamical wind solution for 0.7$M_{\odot}$ star and an X-ray luminosity of 2$\times$10$^{30}$ erg/sec, which is appropriate for TW Hya (e.g. Cleeves et al. 2015). We refer to Ercolano \& Owen (2010, 2016) for specific details about the calculation methods. Assuming an X-ray luminosity of 2.5$\times$10$^{30}$erg/sec we find a flux of $1.07\ \mbox{mJy}$ at $870\ \mu\mbox{m}$ for a distance of $54\ \mbox{pc}$, more than sufficient to explain the observed values. Similar calculations, using a slightly different method and set-up, were performed by Owen, Scaife \& Ercolano (2013), who found a flux of 100$\mu$Jy at 150GHz for a distance of 140 pc, for an X-ray luminosity of 2.5$\times$10$^{30}$erg/sec. Assuming the flux scales as $\nu^{0.6}$, this corresponds to 1.15 mJy at a distance of 54pc, which is in agreement with our new estimates.

Our calculations thus demonstrate that free-free emission from gas, either in a wind or as a bound disk atmosphere can explain the 1mJy flux observed by Andrews et al. (2016).

\subsection{Dust continuum emission}
 If dust is responsible for all of the emission, then following equation 5 of Andrews \& Williams (2005) for a factor five higher temperature, appropriate for the inner $1\ \mbox{au}$ of the disc, we obtain that about $3$ lunar masses of mm-size grains must still be present in the inner disc.
This is in contrast with the calculations of  Alexander \& Armitage (2007), who find significant radial drift, with larger grains disappearing from the inner disc on timescales of $10^4$ yr.
These estimates, however, are tailored to a significantly different disc and photoevaporation model. The mass accretion rate (and thus surface density) assumed in the Alexander \& Armitage (2007) models are about an order of magnitude lower than TW~Hya, and the $\alpha$ viscosity parameter about a factor ten higher than what would be needed to explain the longevity and high mass of the TW~Hya disc ($\sim0.05\ \mbox{M}_{\odot}$ at $10\ \mbox{Myr}$ Bergin et al., 2013; Calvet et al. 2002 suggest $\alpha \sim 10^{-3}$ for a similar mass estimate of $\sim0.06\ \mbox{M}_{\odot}$ at $10\ \mbox{Myr}$)\footnote{Note that CO isotopologue mass measurements are much lower, due to the probable depletion of volatile C in the disc. e.g. Schwarz et al. 2016; Miotello et al. 2016; Kama et al. 2016}.
These two facts alone result in two orders of magnitude difference in the drift timescales, as will be shown below.

An order of magnitude estimate for the disappearance of mm grains from the inner disc of TW~Hya after the photoevaporative gap has opened at $1\ \mbox{au}$ can be obtained as follows. Ignoring factors of order unity and the contribution due to the gas velocity, for small Stokes' numbers ($St<<1$), the radial drift velocity can be expressed as,

\begin{equation}
v_\mathrm{r} = - St \left(\frac{h}{r} \right)^2 v_\mathrm{k},
\end{equation}

\noindent where $h/r$ is the scale hight and $v_k$ is the Keplerian velocity.
The radial drift timescale is thus
\begin{equation}
t_\mathrm{drift}=\frac{R}{|v_\mathrm{r}|} = St^{-1} \left(\frac{h}{r}\right)^{-2} \Omega^{-1}
\end{equation}

The Stokes number of particles with size $s=1\ \mbox{mm}$ is:
\begin{equation}
St = \frac{\pi s \rho_\mathrm{p}}{2\Sigma} = 1.57 \times 10^{-3} \left(\frac{\Sigma}{100\ \mbox{g}/\mbox{cm}^2}\right)^{-1}
\end{equation}

\noindent where $\Sigma$ is the surface density of the gas, and $\rho_\mathrm{p} = 1\ \mbox{g}/\mbox{cm}^3$ is the dust intrisic density.
If the accretion rate is $\sim 10^{-9}\ \mbox{M}_\odot/\mbox{yr}$ at the time of gap opening, then the surface density at $1\ \mbox{au}$ can be obtained from the following,
\begin{equation}
\dot{M}=3 \pi \nu \Sigma = 3 \pi \alpha (h/r)^2 \Omega R^2 \Sigma
\end{equation}
and therefore the drift timescale for the dust from $1\ \mbox{au}$ onto the star, \textbf{with a typical $h/r=0.033$},
 is
\begin{equation}
  t_\mathrm{drift}=\frac{2 \dot{M}}{3\pi^2\nu s \rho_p \Omega}\left(\frac{h}{r}\right)^{-2} = 1.8 \times 10^5\ \mbox{yr}
\end{equation}
This values carries however significant uncertainties, mainly due to its strong dependence on the gas radial velocity, which we only roughly estimate from eq.~1,
and our neglect of the changes in the draining inner disk surface density, which affects the Stokes number.

In brief, dust will remain longer in the inner disc if the viscosity is low -- which is probably the case in the long-lived TW~Hya disc -- and/or if the accretion rate at the time of gap opening is high -- again a probable scenario in TW~Hya given the strong X-ray flux from the central object which results in high photoevaporation rates.
This lends credence to the fact that we might indeed be able to observe the gap opening phase in the TW~Hya disc.

\section{Gas and Dust surface density Evolution under viscosity and X-ray photoevaporation}

We have developed a simple 1D viscous evolution model which includes X-ray photoevaporation for a TW~Hya-like system.

Our 1D gas surface density evolution code, {\rm Spock}, is described in detail in Ercolano \& Rosotti (2015).
At $t=0$ the disc has a mass of $0.15\ \mbox{M}_\odot$, a radius of $18\ \mbox{au}$ and an accretion rate of $10^{-8}\ \mbox{M}_\odot/\mbox{yr}$.
We solve the viscous evolution equation, taking into account the photo-evaporative mass-loss, on a grid of $2500$ points uniformly spaced in $R^{1/2}$ extending from $0.025\ \mbox{au}$ to $2500\ \mbox{au}$.


In our model the gap opens between $1$ and $2\ \mbox{au}$ at an age of $5.7\ \mbox{Myr}$.
At this point the mass accretion rate onto the star is $3.6 \times 10^{-10}\ M_{\odot}/\mbox{yr}$.
$3.5 \times 10^{5}$ years later ($4.6\%$ of the global disc lifetime and $17\%$ of the transition phase lifetime) the mass accretion rate has already dropped to $8.6 \times 10^{-11}\ \mbox{M}_{\odot}/\mbox{yr}$ and the gap outer radius is out to $6\ \mbox{au}$.
This implies that one expects $4.6\%$ of the global protoplanetary disc population and $17\%$ of transition discs to be observed in this early clearing phase\footnote{When talking about transition discs demographics, that transition discs are a very diverse class of objects.
We refer here to objects that are on the verge on dispersal and have developed a hole as a consequence of photoevaporation, which do not include the class of mm-bright ``transition'' discs, which accrete vigorously and simultaneously have large holes.
It is unclear if the latter are objects undergoing a fast {\it transition} phase at all; they may be more long-lived but rare systems, whose large gaps have been carved by dynamical interaction with a companion (see also discussion by Owen \& Clarke, 2012).}.
These numbers show that a small, but not negligible fraction of young protoplanetary disks in nearby star forming regions should be in this evolutionary phase.

\begin{figure}
\begin{center}
\includegraphics[width=0.47\textwidth]{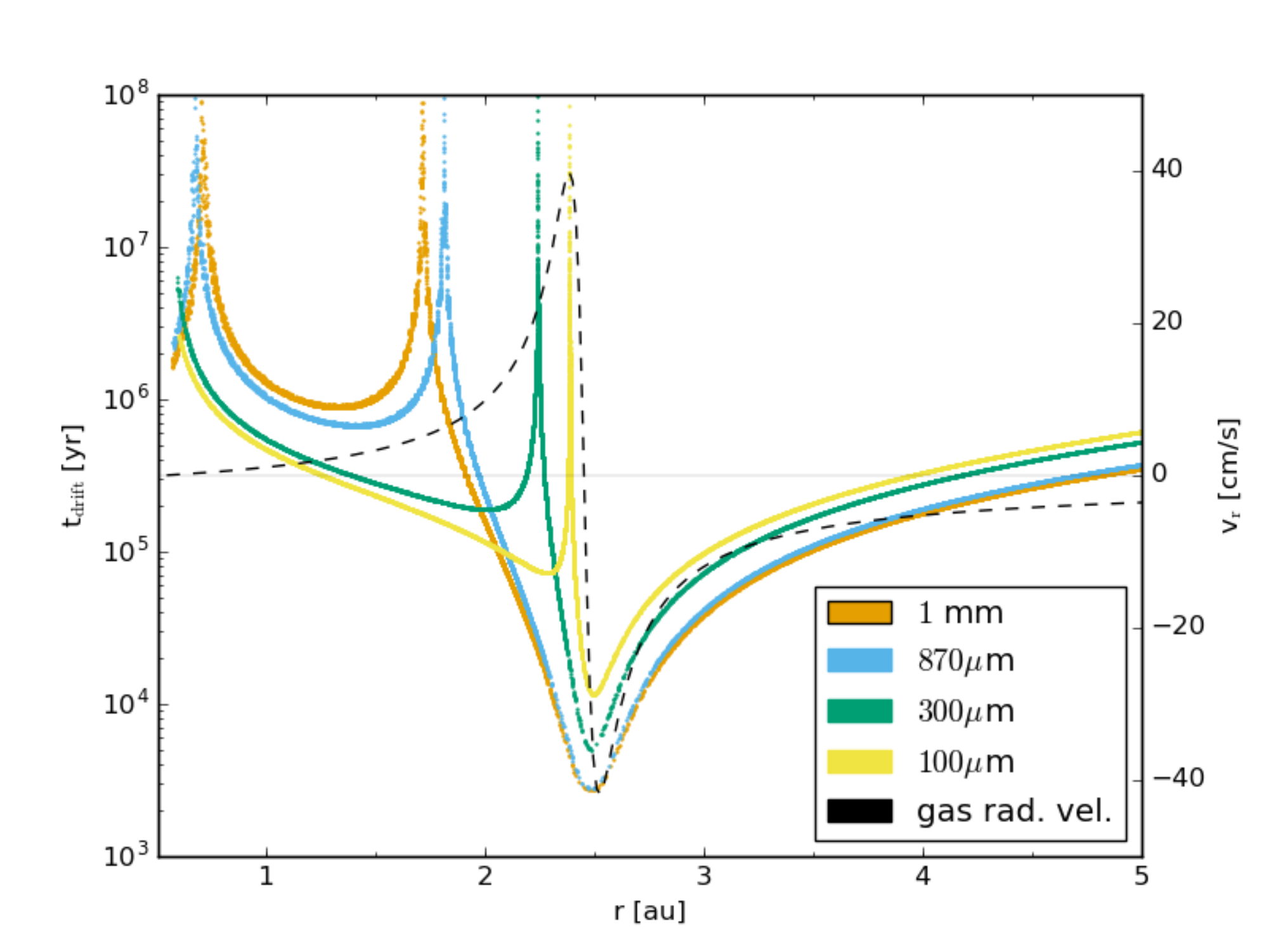}
\caption{The drift timescales of the dust particles of different sizes are plotted at the location and at the time of gap opening.
The radial velocity of the gas is overplotted with a dashed line and its zero value is marked with a grey line.}
\end{center}
\end{figure}

\begin{figure}
\begin{center}
\includegraphics[width=0.47\textwidth]{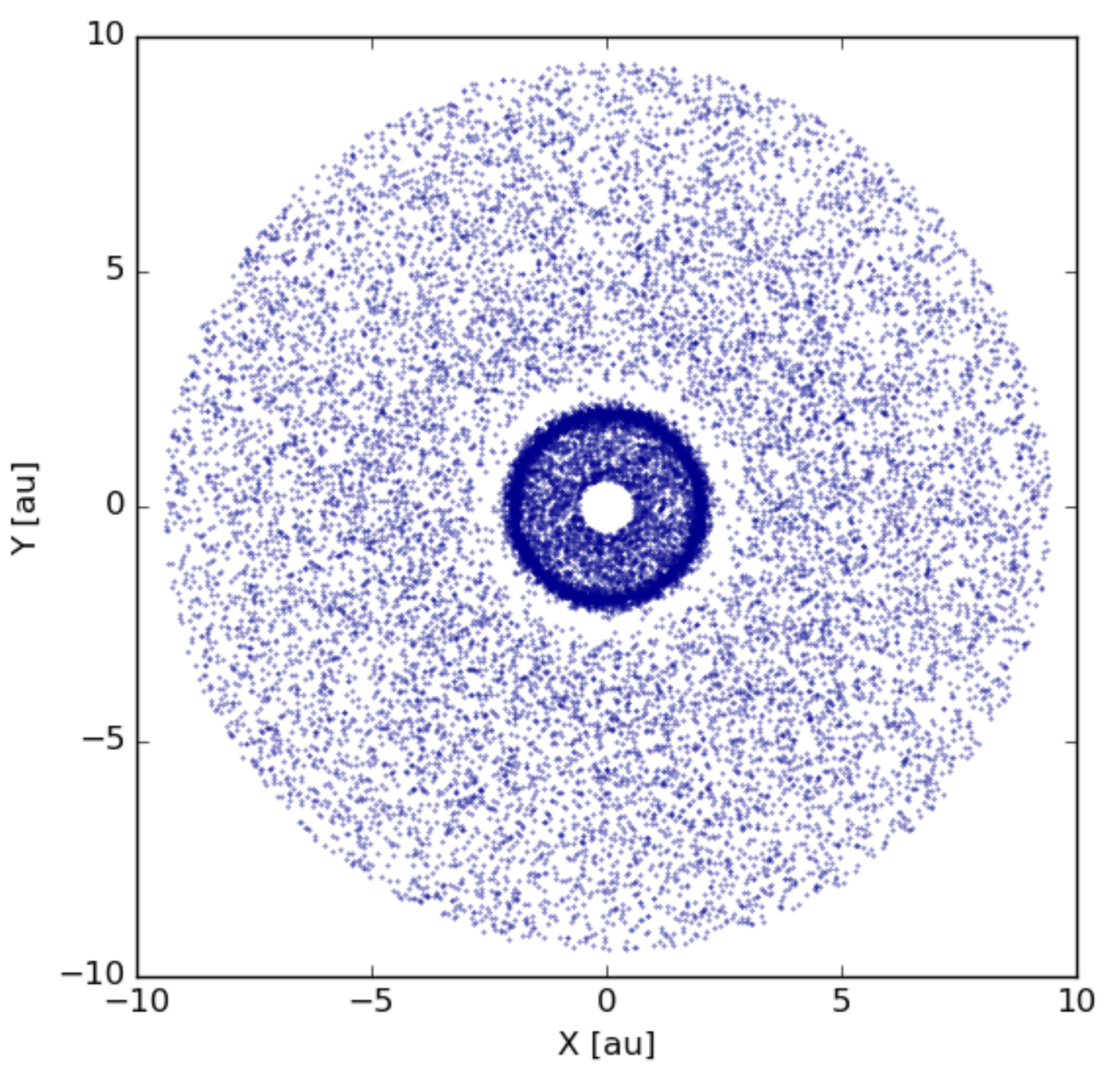}
\caption{The 2D dust distribution for $870\ \mu\mbox{m}$ size particles is shown at the time of gap opening.}
\end{center}
\end{figure}

In order to verify the timescales for dispersal of dust particles of various sizes from the inner disc once the gap is opened, we have used the gas surface density distributions obtained with Spock as an input to the \textsc{fargo} hydro code (Masset 2000) modified by Picogna \& Kley (2015). We model the dust as individual particles, sampling the evolution of a larger ensemble of single dust particles and their interaction with the surrounding environment.

In Figure~1 we show the drift timescales for particles of sizes between $100\ \mu\mbox{m}$ and $1\ \mbox{mm}$ just after the opening of the gap in the dust distribution (as it is shown in the 2D dust distribution for $870\ \mu\mbox{m}$ size particles in Figure~2), when the gas mass accretion rates is $3.7 \times 10^{-10}\ \mbox{M}_{\odot}/\mbox{yr}$. In the inner regions ($\sim 0.5\ \mbox{au}$) particles of $870\ \mu\mbox{m}$ have a drift timescale of $2 \times 10^{6}\ \mbox{yr}$.

The drift timescales have an abrupt change in their distribution inside $2.5\ \mbox{au}$ because the strong gas radial velocity (overplotted with a dashed line) is directed outwards and drags with it the dust particles. Inside $1\ \mbox{au}$ the gas radial velocity is not strong enough and the bigger dust particles ($870 \mu\mbox{m}$, $1\ {mm}$) are able to migrate inwards but with a long drift time-scale.

It is thus likely that some of the $870\ \mu\mbox{m}$ size particles still linger in the inner gas disk of TW Hya and contribute to the emission of the unresolved $1\ \mbox{mJy}$ flux detected by Andrews et al. (2016).

\section{Conclusions}

In this Letter we presented argument in support of X-ray photoevaporation as the origin of the recently detected gap in the continuum emission of the TW Hya disc. In this scenario, TW Hya is an object on the edge of dispersal, where the mass-loss-rate due to X-ray photoevaporation ($\sim$2$\times$10$^{-8} M_{\odot}$, corresponding to the observed X-ray luminosity of TW Hya) is about ten times larger than the (measured) gas accretion rate, and hence in the process of opening a gap between $1$ and $2\ \mbox{au}$. It is not surprising to find an object in this phase, since current statistics show that one should expect of the order of 4 to 6 early phase clearing objects in each of the nearby star forming regions.

We further show that the detection of $\sim$1mJy continuum emission at 870$\mu$m (Andrews et al., 2016) is not in contrast with this scenario. We consider possible sources for the emission, finding that both free-free emission and dust continuum emission, or a combination of the two, are able to explain the observed flux level. Our new photoionisation models, in agreement with previous estimates by Pascucci et al. (2012) and Owen, Scaife \& Ercolano (2013) indicate that free-free emission is produced in the ionised bound disc atmosphere in the draining inner ($< 0.5\ \mbox{au}$) disc.
Sub-mm size dust grains may also contribute to the emission, as they linger in the inner disc thanks to long stopping times in the disc of TW Hya. Indeed we calculate that the stopping time for $870\ \mu\mbox{m}$ grains is about $2\ \mbox{Myr}$ at $0.5\ \mbox{au}$ at the time of gap opening .

In conclusion, we have shown that the X-ray photoevaporation model is consistent with a wide variety of observtional characteristics of the TW Hya disk and, as such, it is the most likely explanation of the observed gap at $1\ \mbox{au}$, instead of the presence of a planet. Our study thus suggests that our closest protoplanetary disc is an object caught at the start of the dispersal phase, making it the best suited natural laboratory to study this phenomenon.


\vspace{-0.4cm}
\section*{Acknowledgements}
We thank the anonymous referee for a constructive report. 
GR has been supported by the DISCSIM project, grant agreement 341137 funded by the European Research Council under
ERC-2013-ADG. LT was partly supported by the Gothenburg Centre for Advanced Studies in Science and Technology as part of the GoCAS program Origins of Habitable Planets and by the Italian Ministero dell'Istruzione, Universita' e Ricerca through the grant Progetti Premiali 2012-iALMA (CUP C52I13000140001.
\bibliographystyle{mnras}

{}

\label{lastpage}

\end{document}